\documentclass{aastex701}

\newcommand{\sunriseiii}{{\sc Sunrise~iii}}
\usepackage{subfigure}
\usepackage{color}
\usepackage{booktabs}
\usepackage{amsmath,amssymb,amsfonts}
\usepackage{times}
\usepackage{here}
\usepackage{multirow}

\received{2026 June 18}
\revised{2026 July 24}
\accepted{2026 July 29}

\begin{document}

\title{Three-dimensional Magnetic Structures of Ellerman Bombs revealed by \sunriseiii/SCIP}

\author[orcid=0000-0001-7452-0656,sname='Kawabata']{Yusuke~Kawabata} \affiliation{National Astronomical Observatory of Japan, 2-21-1 Osawa, Mitaka, Tokyo 181-8588, Japan}\email{kawabata.yusuke@nao.ac.jp}	
\author[orcid=0000-0002-5054-8782,sname='Katsukawa']{Yukio~Katsukawa} \affiliation{National Astronomical Observatory of Japan, 2-21-1 Osawa, Mitaka, Tokyo 181-8588, Japan}\affiliation{Department of Astronomy, The University of Tokyo, 7-3-1, Hongo, Bunkyo-ku, Tokyo 113-0033, Japan}\affiliation{Department of Astronomical Science, The Graduate University for Advanced Studies (SOKENDAI), 2-21-1 Osawa, Mitaka, Tokyo 181-8588, Japan}\email{yukio.katsukawa@nao.ac.jp}	
\author[orcid=0000-0001-5616-2808,sname='Kubo']{Masahito~Kubo} \affiliation{National Astronomical Observatory of Japan, 2-21-1 Osawa, Mitaka, Tokyo 181-8588, Japan}\email{masahito.kubo@nao.ac.jp}	
\author[orcid=0000-0002-7044-6281,sname='Oba']{Takayoshi~Oba} \affiliation{Advanced Research Center for Space Science and Technology, Institute of Science and Engineering, Kanazawa University, Kakuma-machi, Kanazawa, Ishikawa 920-1192, Japan}\affiliation{Max-Planck-Institut für Sonnensystemforschung, Justus-von-Liebig-Weg 3, 37077 Göttingen, Germany}\email{oba-takayoshi@staff.kanazawa-u.ac.jp}	
\author[orcid=0000-0002-1043-9944,sname='Matsumoto']{Takuma~Matsumoto} \affiliation{Centre for Integrated Data Science, Institute for Space-Earth Environmental Research, Nagoya University, Furocho, Chikusa-ku, Nagoya, Aichi 464-8601, Japan}\email{takuma.matsumoto@gmail.com}	
\author[orcid=0000-0002-4669-5376,sname='Ishikawa']{Ryohtaroh~T.~Ishikawa} \affiliation{National Institute for Fusion Science, 322-6 Oroshi-cho, Toki City 509-5292, Japan}\email{ishikawa.ryohtaro@nifs.ac.jp}		
\author[orcid=0000-0001-6793-8528, sname='Naito']{Yoshihiro Naito} 
\affiliation{Department of Astronomical Science, The Graduate University for Advanced Studies (SOKENDAI), 2-21-1 Osawa, Mitaka, Tokyo 181-8588, Japan}
\email{yoshihiro.naito@grad.nao.ac.jp}
\author[orcid=0000-0001-5686-3081,sname='Hara']{Hirohisa~Hara} \affiliation{National Astronomical Observatory of Japan, 2-21-1 Osawa, Mitaka, Tokyo 181-8588, Japan}\email{hirohisa.hara@nao.ac.jp}	
\author[orcid=0000-0003-4764-6856,sname='Shimizu']{Toshifumi~Shimizu} \affiliation{Department of Earth and Planetary Science, The University of Tokyo, 7-3-1, Hongo, Bunkyo-ku, Tokyo 113-0033, Japan}\affiliation{Institute of Space and Astronautical Science, Japan Aerospace Exploration Agency, 3-1-1, Yoshinodai, Chuo-ku, Sagamihara, Kanagawa 252-5210, Japan}\email{shimizu.toshifumi@isas.jaxa.jp}	
\author[orcid=0009-0005-9709-8431,sname='Uraguchi']{Fumihiro Uraguchi}
 \affiliation{National Astronomical Observatory of Japan, 2-21-1 Osawa, Mitaka, Tokyo 181-8588, Japan}
\email{fumihiro.uraguchi@nao.ac.jp}  
\author[orcid=0000-0002-8342-8314,sname='Tsuzuki']{Toshihiro Tsuzuki}
 \affiliation{National Astronomical Observatory of Japan, 2-21-1 Osawa, Mitaka, Tokyo 181-8588, Japan}
\email{toshihiro.tsuzuki@nao.ac.jp}  
\author[sname='Shinoda']{Kazuya Shinoda}
 \affiliation{National Astronomical Observatory of Japan, 2-21-1 Osawa, Mitaka, Tokyo 181-8588, Japan}
\email{shinoda.kazuya@nao.ac.jp}  
\author[sname='Tamura']{Tomonori Tamura}
 \affiliation{National Astronomical Observatory of Japan, 2-21-1 Osawa, Mitaka, Tokyo 181-8588, Japan}
\email{tomonori.tamura@nao.ac.jp}  
\author[orcid=0000-0003-4452-858X,sname='Suematsu']{Yoshinori Suematsu}
 \affiliation{National Astronomical Observatory of Japan, 2-21-1 Osawa, Mitaka, Tokyo 181-8588, Japan}
\email{yoshinori.suematsu@nao.ac.jp} 
\author[orcid=0000-0001-5518-8782,sname='Quintero Noda']{Carlos Quintero Noda} \affiliation{Instituto de Astrofísica de Canarias, Vía Láctea, s/n, E-38205 La Laguna, Spain}
\affiliation{Departamento de Astrofísica, Univ. de La Laguna, La Laguna, Tenerife, E-38200, Spain}
\email{carlos.quintero@iac.es}  

\author[orcid=0000-0002-3387-026X,sname='del~Toro~Iniesta']{Jose~Carlos~del~Toro~Iniesta} \affiliation{Instituto de Astrofísica de Andalucía, CSIC, Glorieta de la Astronomía s/n, 18008 Granada, Spain}\affiliation{Spanish Space Solar Physics Consortium}\email{jti@iaa.es}		
\author[orcid=0000-0001-8829-1938,sname='Orozco~Suárez']{David~Orozco~Suárez} \affiliation{Instituto de Astrofísica de Andalucía, CSIC, Glorieta de la Astronomía s/n, 18008 Granada, Spain}\affiliation{Spanish Space Solar Physics Consortium}\email{orozco@iaa.es}		
\author[sname='Balaguer Jimenez']{María Balaguer Jimenez} \affiliation{Instituto de Astrofísica de Andalucía, CSIC, Glorieta de la Astronomía s/n, 18008 Granada, Spain} \email{balaguer@iaa.es}  

\author[orcid=0000-0002-3418-8449,sname='Solanki']{Sami~K.~Solanki} \affiliation{Max-Planck-Institut für Sonnensystemforschung, Justus-von-Liebig-Weg 3, 37077 Göttingen, Germany}\email{solanki@mps.mpg.de}		
		
\author[orcid=0000-0003-1459-7074,sname='Lagg']{Andreas~Lagg} \affiliation{Max-Planck-Institut für Sonnensystemforschung, Justus-von-Liebig-Weg 3, 37077 Göttingen, Germany}\email{lagg@mps.mpg.de}		
\author[orcid=0000-0002-9972-9840,sname='Gandorfer']{Achim~Gandorfer} \affiliation{Max-Planck-Institut für Sonnensystemforschung, Justus-von-Liebig-Weg 3, 37077 Göttingen, Germany}\email{gandorfer@mps.mpg.de}		

\author[orcid=0000-0002-0787-8954,sname='Bernasconi']{Pietro~Bernasconi} \affiliation{Johns Hopkins University Applied Physics Laboratory, 11100 Johns Hopkins Road, Laurel, Maryland, USA}\email{pietro.bernasconi@jhuapl.edu}		
\author[sname='Berkefeld']{Thomas~Berkefeld} \affiliation{Institut für Sonnenphysik (KIS), Georges-Köhler-Allee 401a, 79110 Freiburg, Germany}\email{thomas.berkefeld@leibniz-kis.de}		
\author[orcid=0009-0009-4425-599X,sname='Feller']{Alex~Feller} \affiliation{Max-Planck-Institut für Sonnensystemforschung, Justus-von-Liebig-Weg 3, 37077 Göttingen, Germany}\email{feller@mps.mpg.de}		
\author[orcid=0000-0001-6317-4380,sname='Riethmüller']{Tino~L.~Riethmüller} \affiliation{Max-Planck-Institut für Sonnensystemforschung, Justus-von-Liebig-Weg 3, 37077 Göttingen, Germany}\email{riethmueller@mps.mpg.de}	

\author[orcid=0000-0001-9228-3412,sname='Álvarez-Herrero']{Alberto~Álvarez-Herrero} \affiliation{Instituto Nacional de T\'ecnica Aeroespacial (INTA), Ctra. de Ajalvir, km. 4, E-28850 Torrejón de Ardoz, Spain}\affiliation{Spanish Space Solar Physics Consortium}\email{alvareza@inta.es}		
\author[orcid=0000-0003-3490-6532,sname='Smitha']{H.~N.~Smitha} \affiliation{Max-Planck-Institut für Sonnensystemforschung, Justus-von-Liebig-Weg 3, 37077 Göttingen, Germany}\email{narayanamurthy@mps.mpg.de}		
\author[sname='Grauf']{Bianca~Grauf} \affiliation{Max-Planck-Institut für Sonnensystemforschung, Justus-von-Liebig-Weg 3, 37077 Göttingen, Germany}\email{grauf@mps.mpg.de}		
\author[sname='Carpenter']{Michael~Carpenter} \affiliation{Johns Hopkins University Applied Physics Laboratory, 11100 Johns Hopkins Road, Laurel, Maryland, USA}\email{michael.carpenter@jhuapl.edu}		
\author[sname='Bell']{Alexander~Bell} \affiliation{Institut für Sonnenphysik (KIS), Georges-Köhler-Allee 401a, 79110 Freiburg, Germany}\email{albe@leibniz-kis.de}		
\author[orcid=0000-0001-7764-6895,sname='Martínez~Pillet']{Valentín~Martínez~Pillet} \affiliation{Instituto de Astrofísica de Canarias, Vía Láctea, s/n, E-38205 La Laguna, Spain}\affiliation{Spanish Space Solar Physics Consortium}\email{vmpillet@iac.es}		

\author[orcid=0000-0002-7318-3536,sname='Bailén']{Francisco~Javier~Bailén} \affiliation{Instituto de Astrofísica de Andalucía, CSIC, Glorieta de la Astronomía s/n, 18008 Granada, Spain}\affiliation{Spanish Space Solar Physics Consortium}\email{fbailen@iaa.es}		
\author[orcid=0000-0002-2055-441X,sname='Blanco~Rodríguez']{Julian~Blanco~Rodríguez} \affiliation{Universitat de Valencia Catedrático José Beltrán 2, E-46980 Paterna-Valencia, Spain}\affiliation{Spanish Space Solar Physics Consortium}\email{julian.blanco@uv.es}		
\author[orcid=0000-0003-4319-2009,sname='Castellanos~Durán']{Juan~Sebastián~Castellanos~Durán} \affiliation{Max-Planck-Institut für Sonnensystemforschung, Justus-von-Liebig-Weg 3, 37077 Göttingen, Germany}\email{castellanos@mps.mpg.de}		
\author[orcid=0009-0002-6808-5154,sname='Harnes']{Edvarda~Harnes} \affiliation{Max-Planck-Institut für Sonnensystemforschung, Justus-von-Liebig-Weg 3, 37077 Göttingen, Germany}\email{harnes@mps.mpg.de}		
\author[orcid=0000-0001-6029-7529,sname='Hoelken']{Johannes~Hoelken} \affiliation{Max-Planck-Institut für Sonnensystemforschung, Justus-von-Liebig-Weg 3, 37077 Göttingen, Germany}\email{hoelken@mps.mpg.de}		
\author[orcid=0000-0003-1409-1145,sname='Iglesias']{Francisco~A.~Iglesias} \affiliation{Max-Planck-Institut für Sonnensystemforschung, Justus-von-Liebig-Weg 3, 37077 Göttingen, Germany}\affiliation{Grupo de Estudios en Heliofísica de Mendoza, CONICET, Universidad de Mendoza, Boulogne sur Mer 683, 5500 Mendoza, Argentina}\email{iglesias@mps.mpg.de}		
\author[orcid=0000-0003-0175-6232,sname='Siu-Tapia']{Azaymi~L.~Siu-Tapia} \affiliation{Instituto de Astrofísica de Andalucía, CSIC, Glorieta de la Astronomía s/n, 18008 Granada, Spain}\affiliation{Spanish Space Solar Physics Consortium}\email{siu@iaa.es}		
\author[orcid=0000-0003-1483-4535,sname='Strecker']{Hanna~Strecker} \affiliation{Instituto de Astrofísica de Andalucía, CSIC, Glorieta de la Astronomía s/n, 18008 Granada, Spain}\affiliation{Spanish Space Solar Physics Consortium}\email{streckerh@iaa.es}		
\author[orcid=0000-0003-1971-5551,sname='Vukadinović']{Dušan~Vukadinović} \affiliation{Institut für Physik, Universität Graz, Universitätsplatz 5, 8010 Graz, Austria}\affiliation{Max-Planck-Institut für Sonnensystemforschung, Justus-von-Liebig-Weg 3, 37077 Göttingen, Germany}\email{vukadinovic@mps.mpg.de}	
\begin{abstract}
Ellerman bombs (EBs) are widely recognized as photospheric and chromospheric signatures of magnetic reconnection.
 However, the three-dimensional (3D) magnetic topology has remained elusive due to the lack of seamless height coverage in observations. 
 Here, we present initial results from the \sunriseiii/SCIP (Sunrise Chromospheric Infrared spectroPolarimeter) observations of an emerging flux region. 
 Exploiting the seeing-free, high-spatial-resolution observations provided by the 1-meter balloon-borne telescope, SCIP achieved seamless multi-line spectropolarimetry from the photosphere to the lower chromosphere.
 We analyzed the multi-line Stokes profiles of the photospheric Fe \textsc{i} and K \textsc{i} lines and the chromospheric Ca \textsc{ii} lines, and applied the Weak Field Approximation to the K \textsc{i} and Ca \textsc{ii} lines to reconstruct the 3D magnetic field structure. 
 The blue- and red-wing brightenings of the Ca \textsc{ii} 8542 \AA\ line appear at spatially offset locations, indicating bi-directional reconnection flows. The reconstructed 3D magnetic field reveals that the opposite-polarity field structure reaches different heights in the two events analyzed. 
 In one event, it is confined to the lower layers and is absent at the formation height of Ca \textsc{ii} 8542 \AA\ core, which shows no intensity enhancement, whereas in the other event it extends up to the Ca \textsc{ii} 8542 \AA\ core formation height, where enhanced line-core intensity is also observed. We interpret this as the reconnection current sheet reaching different altitudes. These results demonstrate that SCIP has successfully resolved the 3D structure of EBs, distinguishing magnetic reconnection events occurring at different atmospheric heights.
\end{abstract}


\section{Introduction} 
Magnetic reconnection in the solar atmosphere is a fundamental process that converts magnetic energy into thermal and kinetic energy. 
In the lower atmosphere, specifically within the photosphere and low chromosphere, this process manifests as Ellerman bombs (EBs). 
First reported by \citet{1917ApJ....46..298E}, EBs are characterized by transient brightenings in the wings of H$\alpha$ while no significant increase is observed in the line core intensity \citep{1983SoPh...87..135K, 2002ApJ...575..506G, 2011ApJ...736...71W, 2013ApJ...774...32V, 2015ApJ...812...11V}.
EBs are ubiquitous phenomena, observed not only in active regions \citep{2013JPhCS.440a2007R}, particularly in emerging flux regions, but also in the quiet Sun \citep{2016A&A...592A.100R, 2020A&A...641L...5J}.

The study of EBs is crucial for understanding the physics of magnetic reconnection in a unique plasma regime distinct from the solar corona.
The lower solar atmosphere is characterized by a moderate plasma $\beta$ (the ratio of gas pressure to magnetic pressure) \citep{2001SoPh..203...71G} and is dominated by neutral atoms, forming a weakly ionized plasma environment \citep{1993ApJ...406..319F}.
Theoretical studies suggest that collisional effects between ions and neutrals in such partially ionized plasmas significantly alter magnetic diffusivity and the efficiency of energy conversion during reconnection \citep{2008A&A...486..569S}. 
Furthermore, the plasma $\beta$ also plays a crucial role in regulating the reconnection efficiency, with different beta regimes leading to distinct dynamics \citep{2019A&A...628A...8P}.
Therefore, diagnosing the magnetic topology and velocity field at the precise height of energy release is essential to understand how reconnection proceeds in this specific environment.

However, observational verification of the three-dimensional (3D) magnetic structure of EBs has been challenging. Obtaining simultaneous, high-resolution measurements of magnetic fields from the photosphere to the chromosphere is required to trace the roots of reconnection.
In particular, it is important to access multiple spectral lines spanning a continuous range of formation heights from the photosphere through the upper photosphere to the lower chromosphere; see, e.g., \citet{2024ApJ...960...26K} for the corresponding formation height stratification of the lines used in this work.
To address these challenges, the Sunrise Chromospheric Infrared spectroPolarimeter \citep[SCIP;][]{2020SPIE11447E..0YK, 2026SoPh..301...99K} offers an unique capability.
SCIP was on board the \sunriseiii\ balloon-borne observatory, which is the third flight of the  SUNRISE project  \citep{2010ApJ...723L.127S, 2011SoPh..268....1B, 2017ApJS..229....2S, 2025SoPh..300...75K}; see also \cite{2026arXiv260607989S} for the science overview of \sunriseiii.
\sunriseiii\ provides stable, high-spatial-resolution observations unaffected by atmospheric seeing.
SCIP performs simultaneous spectropolarimetry in multiple lines covering a wide range of formation heights.
 Synthetic observations based on magnetohydrodynamic (MHD) simulations have demonstrated that this combination of lines allows for the retrieval of seamless information on magnetic fields and velocities from the photosphere to the chromosphere, enabling the distinction between low- and high-altitude reconnection events \citep{2024ApJ...960...26K}.

In this paper, we present initial results from the \sunriseiii/SCIP observations of an emerging flux region obtained during its successful flight in July 2024. 
We focus on transient brightenings observed in the Ca~{\sc ii} wings, corresponding to EBs. By utilizing the Weak Field Approximation (WFA), we investigate the magnetic topology associated with these events. 
We report the detection of bi-directional flows and three-dimensional structure of anti-parallel magnetic fields, demonstrating the power of multi-line spectropolarimetry in resolving the complex nature of low-atmosphere reconnection.

\section{Observations}
The observations used in this work were carried out during the flight of the \sunriseiii\ balloon-borne solar observatory in July 2024 (observing program 29\_EMEF; see Table~1 of \cite{2026arXiv260607989S}).
We utilized SCIP, which allows for simultaneous spectropolarimetric observations of the photosphere and chromosphere with a designed spatial resolution of about $0.2''$.
The 1-meter aperture of \sunriseiii\ and the absence of atmospheric seeing distortions enable the detection of fine-scale magnetic structures.

The target of this study was an emerging flux region that had not yet been assigned a NOAA active-region number at the time of the SCIP observations on 2024 July 15; it was subsequently catalogued as NOAA AR~13753 on the following day. The observed region was centered at helioprojective coordinates $(x, y) \approx (235'', 120'')$ at the beginning of the observing sequence and drifted to $\approx (265'', 122'')$ by its end, following the solar rotation of the target region, which corresponds to a heliocentric angle of $\mu \approx 0.95$--$0.96$.
Continuous observations were successfully performed for approximately 4 hours, from 10:05:01 UT to 14:12:27 UT on 2024 July 15. Flux emergence had been active in this region for several hours prior to the SCIP observations and continued throughout the observing sequence.
The high stability of the observations was largely attributed to the performance of the pointing system of the gondola \citep{2025SoPh..300..112B}, and the Correlation Wave-front Sensor \citep{2026SoPh..301...57B}.
The field of view was approximately $58'' \times 58''$.
The observed spectral lines include Fe~{\sc i} 8468\,\AA\ and the K~{\sc i} D$_{1}$/D$_{2}$ lines (7698\,\AA\ and 7664\,\AA) probing the photosphere, the Ca~{\sc ii} 8498\,\AA\ and the Ca~{\sc ii} 8542\,\AA\  lines probing the lower chromosphere, although the Ca~{\sc ii} 8498\,\AA\ line is sensitive to a slightly lower atmospheric layer compared to the 8542 \AA\ line \citep{2017MNRAS.464.4534Q}.
This multi-line configuration allows for a seamless diagnosis of the magnetic field stratification from the photosphere to the chromosphere \citep{2016MNRAS.459.3363Q, 2017MNRAS.464.4534Q, 2017MNRAS.472..727Q, 2019MNRAS.486.4203Q, 2023MNRAS.523..974M, 2024ApJ...960...26K}.

The instrument operated in a scanning mode, acquiring spectral slit images to construct 2D maps.
Each 2D map was obtained in a single scan of 11 minutes and 56 seconds, with an accumulation time of 1 second per slit position. During the 4-hour observing window, 21 consecutive scans were performed. Because the last scan was interrupted before completion, the full field of view was covered in the first 20 scans, yielding 20 complete 2D maps of the target region.

\section{Data Reduction and Analysis}\label{sec:datared}
We performed data reduction steps including dark and flat-field corrections, skew correction, and wavelength and polarimetric calibrations.
The resulting spectral sampling corresponds to 39.5 mÅ pixel$^{-1}$ in the 8500 \AA \ window and 36 mÅ pixel$^{-1}$ in the 7700 \AA \ window.
The spatial sampling is $0.094''$ pixel$^{-1}$.
The noise level for Stokes $Q/I_c$, $U/I_c$, and $V/I_c$ is estimated to be $1 \times 10^{-3}$ at the $1\sigma$ level, where $I_c$ is the continuum intensity.

To infer the line-of-sight (LOS) magnetic field component ($B_{\mathrm{LOS}}$) from the observed Stokes profiles, we applied the WFA method \citep{1956PASJ....8..108U, 1989ApJ...343..920J} to the K \textsc{i} and Ca \textsc{ii} lines, adopting the spatially regularized implementation of \cite{2020A&A...642A.210M}.
This approximation is valid when the Zeeman splitting $\Delta \lambda_B$ is significantly smaller than the Doppler width $\Delta \lambda_D$ of the spectral line. 
Under this condition, the Stokes $V$ profile is proportional to the wavelength derivative of the Stokes $I$ intensity profile. The relationship is described as follows:
\begin{equation}
    V(\lambda) = -\Delta \lambda_B f \bar{g} \cos \theta \frac{\partial I}{\partial \lambda} ,
       \label{eq:wfa}
\end{equation}
where $\Delta \lambda_B$ is the Zeeman splitting, $f$ is the magnetic filling factor (assumed to be unity), $\bar{g}$ is the effective Land\'{e} factor, $\theta$ is the angle between the magnetic field vector and the line of sight.

For the K \textsc{i} 7664 and 7698 \AA\ lines, we adopted effective Landé factors ($\bar{g}$) of 1.16 and 1.33, respectively, with a wavelength range of $\pm 0.20$\,\AA \ from the line center. For Ca \textsc{ii} 8498 \AA\ ($\bar{g}=1.07$), the range was set to $\pm 0.45$ \AA. In the case of Ca \textsc{ii} 8542 \AA\ ($\bar{g}=1.10$), we used $\pm 0.5$ \AA\ for the core and wavelength windows symmetrically offset from the line center by 1.5--2.0 \AA\ (i.e., $[\lambda_0-2.0,\lambda_0-1.5]$ \AA\ and $[\lambda_0+1.5,\lambda_0+2.0]$ \AA) for the wings, which are formed in the photosphere.
 Fe~\textsc{i} 8468\,\AA\ was not used for the WFA in this work because the strong magnetic fields in the emerging flux region cause the Zeeman splitting to exceed the Doppler width.
A critical advantage of using multi-line observations with the WFA is the ability to diagnose magnetic fields at different atmospheric heights. 
Previous studies have demonstrated that applying the WFA to the line core of Ca~{\sc ii} 8542\,\AA\ probes the chromospheric magnetic field, whereas applying it to the line wings retrieves information from the photosphere \citep{2018ApJ...866...89C, 2024ApJ...960...26K}.

\section{Results}
\begin{figure}[h]
\centering
\includegraphics[width=\columnwidth,clip]{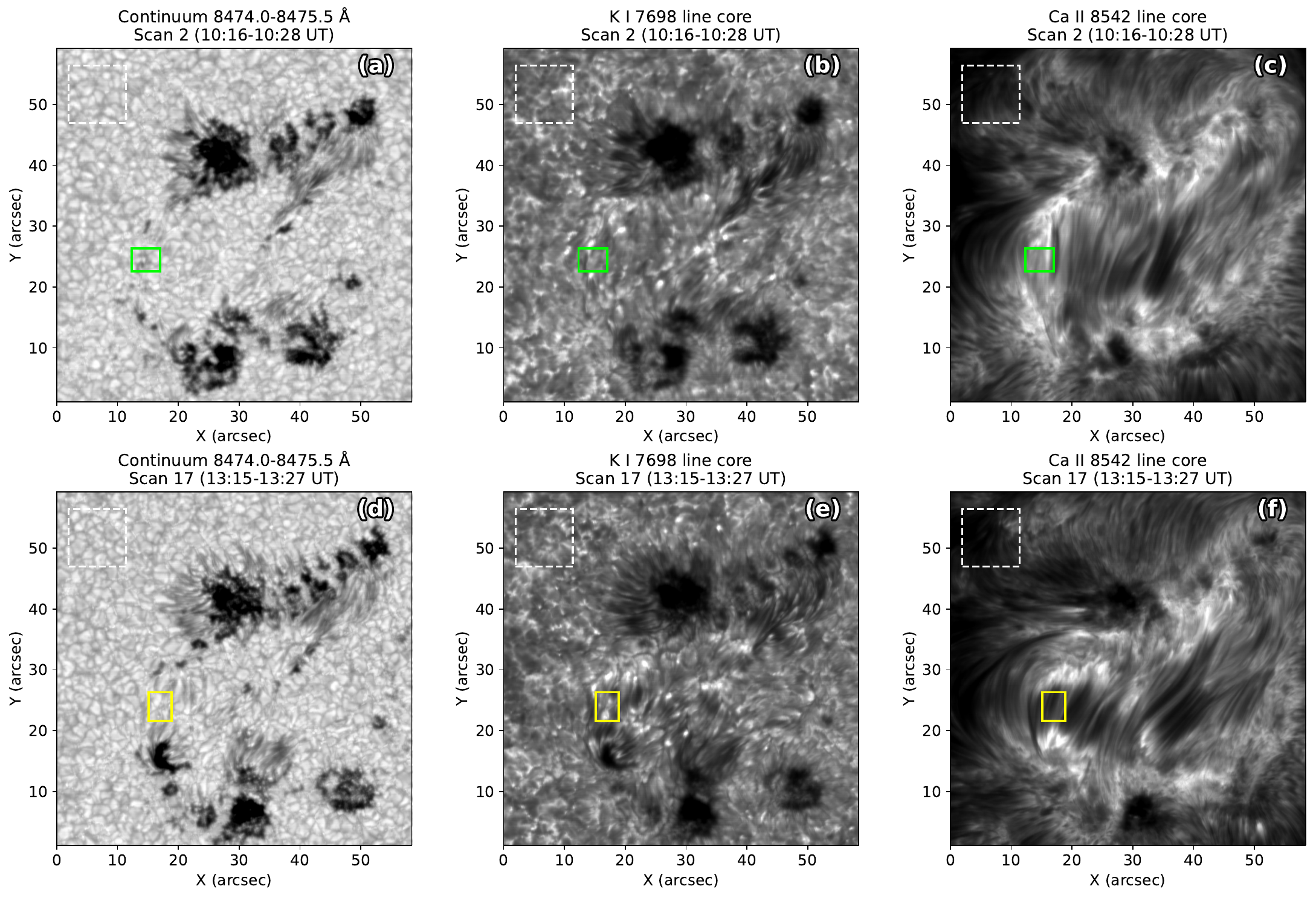}
\caption{Examples of Stokes $I$ maps in the continuum (averaged over 8474.0--8475.5 \AA), the K \textsc{i} 7698 \AA\ line core, and the Ca \textsc{ii} 8542 \AA\ line core, acquired with \sunriseiii/SCIP. Panels (a)--(c) show the 2nd scan (10:17--10:29 UT), in which Event~2 was identified; the green rectangle in each top-row panel marks the field of view of Event~2. Panels (d)--(f) show the 17th scan (13:16--13:28 UT), in which Event~1 was identified; the yellow rectangle in each bottom-row panel marks the field of view of Event~1. The white dashed rectangle (visible near the upper-left of each panel) marks the quiet-Sun reference region used for computing the mean quiet-Sun Stokes $I$ profile. The $x$- and $y$-axes are given in arcsec and correspond to the scanning and slit directions, respectively. An animation of this figure, covering the 20 completed scans (10:05--14:04 UT), is available in the online journal.}
\label{example}
\end{figure}

Figure \ref{example} shows examples of Stokes $I$ scanning maps observed with SCIP. The panels display the spatial distribution of the continuum intensity and the line-core intensity of the K \textsc{i} 7698 \AA\ and Ca \textsc{ii} 8542 \AA\ lines from the two scans (17th and 2nd) in which Event~1 and Event~2 were identified, respectively.
During the 4-hour observation, prominent flux emergence was detected around $(X, Y) \sim (16'', 17'')$ in Figure~\ref{example}, accompanied by numerous brightening events in its vicinity. In this section, we select two of these events (Event 1 and 2) and present the analysis results.
The fields of view of Events 1 and 2, together with the quiet-Sun reference region used to compute the mean quiet-Sun Stokes $I$ profile (see below), are marked in Figure~\ref{example}. A movie covering 20 scans is provided in the online supplementary material.

\begin{figure}[h]
\centering
\includegraphics[width=\columnwidth,clip]{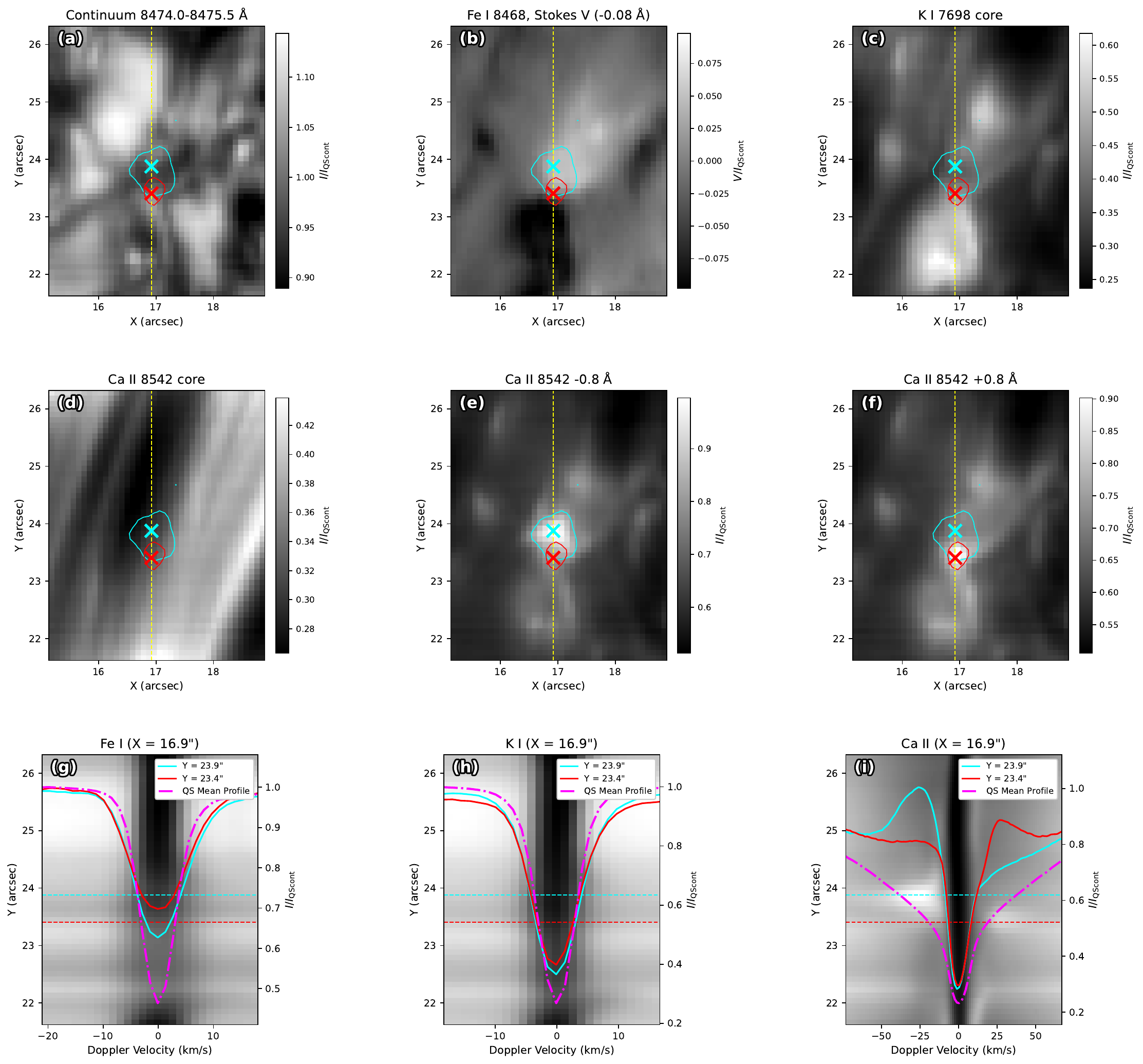}
\caption{Spectral features of Event1. (a) continuum intensity (averaged over 8474.0--8475.5 \AA), (b) Stokes $V$ map of Fe \textsc{i} 8468 \AA\ (at $-0.08$ \AA\ from the line core), and (c) line-core intensity of K \textsc{i} 7698 \AA. (d) Ca \textsc{ii} 8542 \AA\ line-core intensity, and (e) intensities at the blue wing ($-0.8$ \AA) and (f) red wing ($+0.8$ \AA) of the Ca \textsc{ii} line.
The spatial coordinates are given in arcsec and correspond to those of Figure~\ref{example}.
All intensities and the Stokes $V$ map in panels (a)--(f) are normalized by the quiet-Sun continuum, computed from the quiet-Sun reference region defined in Figure~\ref{example}.
The red and cyan contours indicate the regions of intensity enhancement in the Ca \textsc{ii} red wing and Ca \textsc{ii} blue wing, respectively.
The yellow dashed line indicates the position for the wavelength–space plots shown in the panels (g)-(i).
The wavelength scale is converted to Doppler velocity, which is defined relative to the mean line centroid of the quiet-Sun reference region within the same dataset (the white dashed rectangle in Figure~\ref{example}), i.e., taken as $0\ \mathrm{km\,s^{-1}}$.
The cyan and red crosses overplotted on panels (a)--(f) mark the slit positions whose Stokes $I$ profiles are shown by the cyan and red curves in panels (g)--(i), respectively (also normalized by $I_{\mathrm{QScont}}$ of each line).
The magenta lines in the panels (g)-(i) represent the mean Stokes $I$ profiles of the quiet Sun.}
\label{EB1}
\end{figure}

Figure \ref{EB1} displays the spectral features of Event 1 obtained by SCIP. As shown in the panels (e) and (f), brightenings are observed in the Ca \textsc{ii} wings.
As implied by the Stokes $V$ map of Fe \textsc{i} 8468 \AA \ in the panel (b), these brightenings are located between the opposite magnetic polarities.
The panels (g)--(i) display wavelength-space plots along the cut marked in the panels (a)--(f).
In the Fe \textsc{i} and K \textsc{i} lines, the profiles show no significant shift from the average Quiet Sun profile (magenta), while their line core intensities are enhanced. 
In contrast, the Ca \textsc{ii} 8542 Å profile exhibits wing enhancement without significant core brightening compared to the Quiet Sun.
This feature is characteristic of Ellerman bombs. Furthermore, within the Ca~\textsc{ii} wing-brightening region described above, Stokes $I$ profiles showing blue and red asymmetries are observed near the centers of the magnetic polarities (the cyan and red lines in the panel (i)).

\begin{figure}[h]
\centering
\includegraphics[width=\columnwidth,clip]{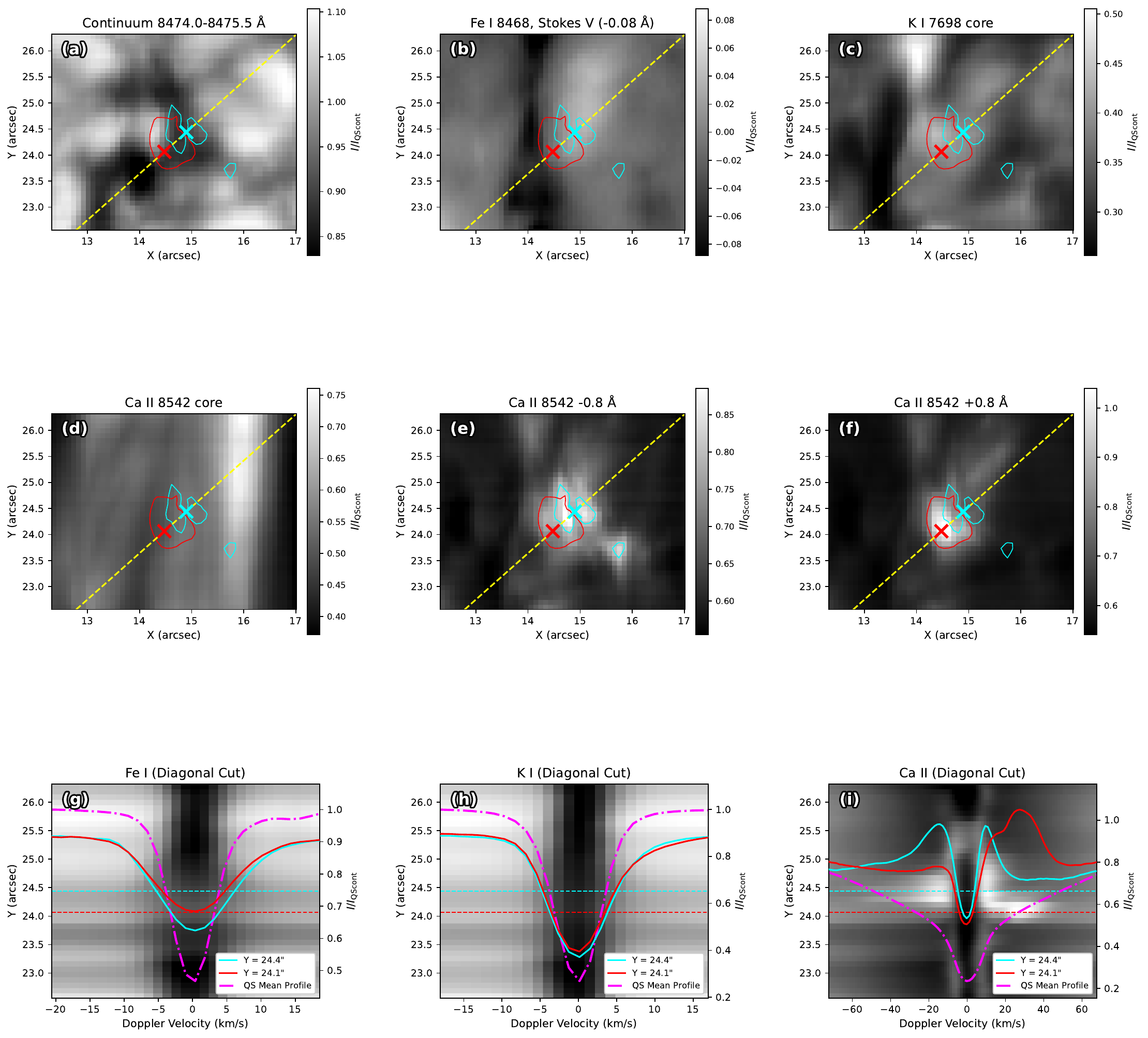}
\caption{The figure format and symbols are the same as in Figure \ref{EB1}, but for a different event (Event 2). In contrast to Figure \ref{EB1}, the wavelength--space plots in panels (g)--(i) are extracted along the inclined yellow dashed line shown in panels (a)--(f). The spectra sampled along this cut are displayed as a function of the $Y$ coordinate of the sampling points; the vertical axis of panels (g)--(i) therefore covers the same $Y$ range as panels (a)--(f), and does not represent the distance along the cut.}
\label{EB2}
\end{figure}

Figure \ref{EB2} displays the spectral features of Event 2 obtained by SCIP.
The figure format and symbols are the same as in Figure \ref{EB1}.
Similar to Event 1, Ca \textsc{ii} wing brightenings are observed between the magnetic polarities seen in the panel (b).
Notably, the continuum maps of Event~2 (panel (a)) show locally darker patches near the brightening site, suggesting that Event~2 occurs above small pore-like structures.
The core enhancement in the photospheric lines (Fe \textsc{i} and K \textsc{i}) also follows the same trend as in Event 1.
However, the Ca \textsc{ii} profile differs from the Event 1 by showing enhancement in both the wings and the line core compared to the Quiet Sun profile.
We note that the line-core intensity around Event~2 (about twice the quiet-Sun level) is not confined to the wing-brightening site as a compact kernel; the surrounding area, which is covered by bright fibrils, shows a comparable core intensity (panel (d)). The interpretation of the core enhancement is discussed in Section 5.
A prominent red asymmetry is observed (red profile in the panel (i)); while the blue-wing brightening has a comparable intensity to the red wing, its tail extends further from the line core compared to the red side (cyan profile in the panel (i)).

\begin{figure}[h]
\centering
\includegraphics[width=\columnwidth,clip]{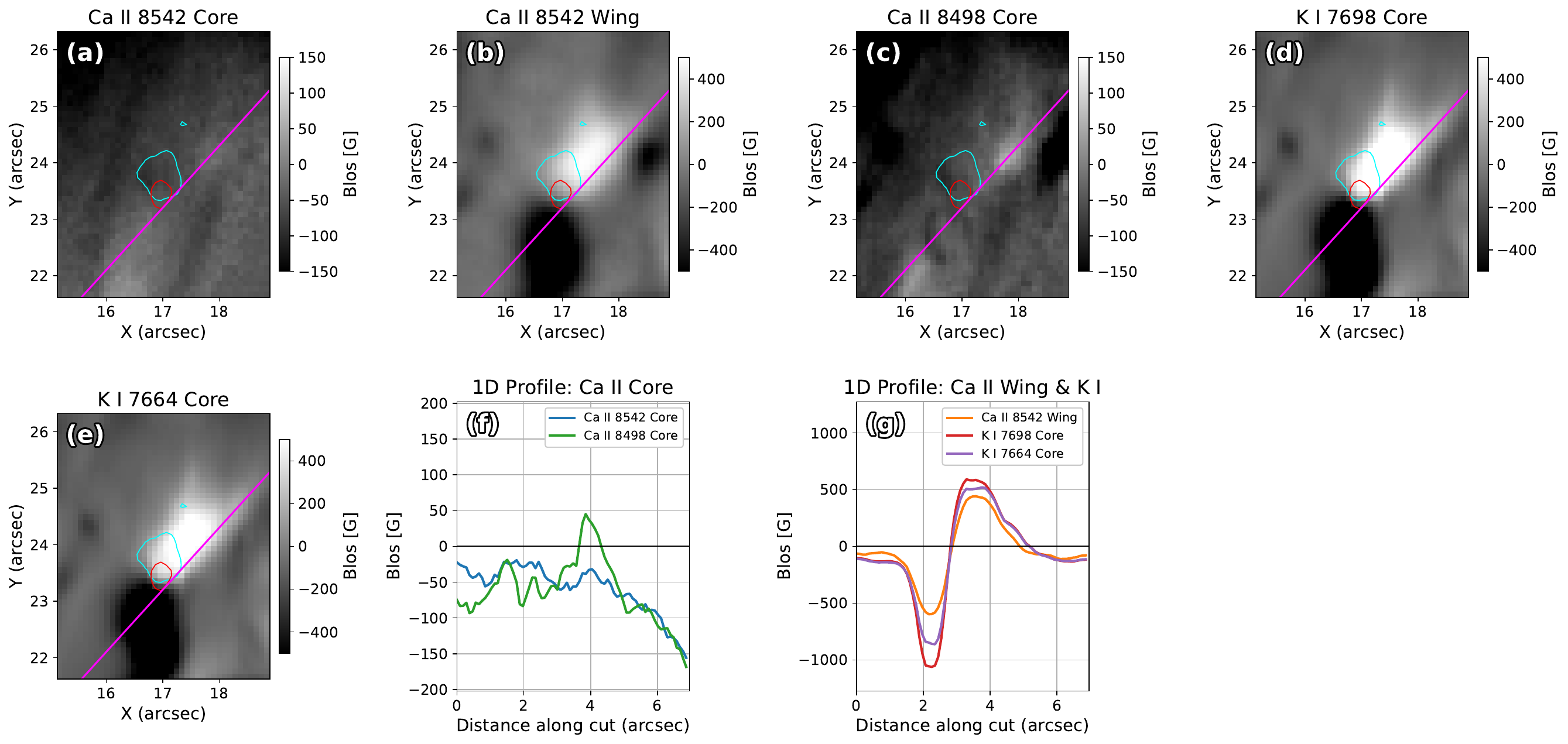}
\caption{Results of the WFA applied to Event 1. The red and cyan contours on the WFA maps are the same as those in Figure~\ref{EB1}, indicating the intensity enhancements in the Ca \textsc{ii} red wing and Ca \textsc{ii} blue wing (panels (e) and (f) of Figure~\ref{EB1}), respectively. The magenta line is oriented so as to cross the two opposite-polarity patches of the photospheric LOS magnetic field (e.g., panel (d)). The one-dimensional spatial profiles of the line-of-sight magnetic field along the magenta line are displayed in panels (f) and (g). The spatial coordinates are given in arcsec and correspond to those of Figure~\ref{example}.}
\label{EB1_WFA}
\end{figure}

Figure \ref{EB1_WFA} presents the results of the WFA applied to the Event 1 region.
The panels (a)--(e) show the 2D maps and the panels (f) and (g) show the 1D profiles along the magenta line.
As also indicated in Figure \ref{EB1}, the Ca \textsc{ii} wing brightening is located at the polarity inversion line of the LOS magnetic field derived from the photospheric lines.
A decreasing trend in the magnetic field strength is observed in the order of K \textsc{i} 7698 \AA, K \textsc{i} 7664 \AA, and the Ca \textsc{ii} 8542 wing. The fields derived from Ca \textsc{ii} 8498 \AA \ and Ca \textsc{ii} 8542 \AA \ are even weaker; while the opposite polarity structure is barely discernible in Ca \textsc{ii} 8498 \AA, the Ca \textsc{ii} 8542 \AA \ map exhibits a completely unipolar field configuration.

\begin{figure}[h]
\centering
\includegraphics[width=\columnwidth,clip]{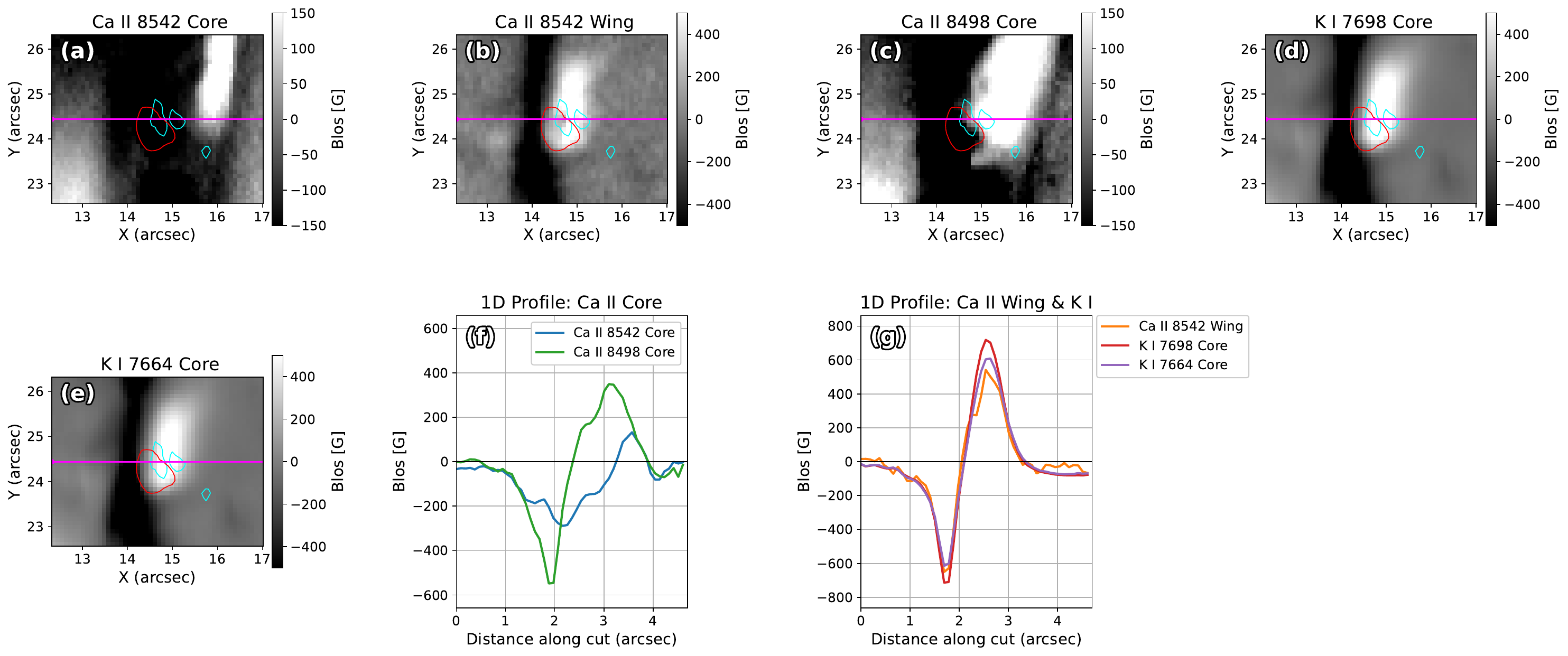}
\caption{Same as Figure \ref{EB1_WFA}, but showing the results of the WFA applied to Event 2.}
\label{EB2_WFA}
\end{figure}

Figure \ref{EB2_WFA} presents the WFA results for Event 2, following the same format as Figure \ref{EB1_WFA}. In contrast to Event 1, the opposite-polarity structure persists up to height of the Ca \textsc{ii} 8542 Å line core, indicating that the bipolar magnetic configuration of Event~2 extends to greater heights than that of Event~1. The decreasing trend of the LOS magnetic field strength from the K lines to the Ca \textsc{ii} wing is consistent with Event 1. However, in the Ca \textsc{ii} core, we observe a spatial shift between the peak intensities of the negative and positive polarities, which we interpret as a height-dependent change in the geometry of the magnetic configuration.

\section{Discussion \& Conclusion}
The blue and red asymmetries in the Ca \textsc{ii} 8542 Å wings are detected at spatially offset locations.
We interpret these features as upflows and downflows of heated plasma, respectively, indicating the presence of bi-directional flows driven by magnetic reconnection.
We emphasize that we use these asymmetries only as qualitative indicators of oppositely directed line-of-sight flows, and we do not translate the Doppler offsets of the enhanced wings into flow velocities. The Ca \textsc{ii} 8542 \AA\ line is formed under optically thick conditions, and the observed profiles are additionally modified by absorption from overlying material, so that the asymmetric enhancements are unlikely to be clean emission peaks. The apparent Doppler offsets therefore do not directly measure the reconnection outflow speed and provide, at best, a rough upper limit on it. A quantitative determination of the velocity field requires NLTE inversions of the full Stokes profiles, which will be addressed in future work.
In contrast, no strong downflows were observed in the photospheric lines, such as Fe \textsc{i} and K \textsc{i}. 
Qualitatively, this is expected because these lines are formed at low heights in the strongly stratified solar atmosphere. 
At these low altitudes, the plasma density is significantly higher, resulting in downflow velocities that are smaller.
In the synthetic SCIP observations of \cite{2024ApJ...960...26K}, which were computed from a radiative MHD simulation of an EB, the Fe \textsc{i} and K \textsc{i} lines exhibited redshifts of several km s$^{-1}$ at the EB site. Such shifts are not detected in our observations. This apparent discrepancy may indicate that the MHD simulation used for the spectral synthesis overestimated the downflow velocities in the photospheric layers around the EB; indeed, previous observations have reported photospheric flows around EBs as small as $\sim$0.2 km s$^{-1}$ \citep{2008PASJ...60...95M}.
Previous studies using Ca \textsc{ii} 8542 \AA \  \citep{2013ApJ...774...32V}, H$\alpha$ \citep{2008PASJ...60...95M, 2011ApJ...736...71W} and Si \textsc{iv} \citep[UV burst;][]{2014Sci...346C.315P,2019A&A...627A.101V, 2020A&A...633A..58O} have reported similar bi-directional signatures. 
The fact that the bi-directional signatures are prominent in the Ca \textsc{ii} wings (formed in the upper photosphere) but less distinct in the K \textsc{i} lines suggests that the reconnection X-point for the observed EBs likely lies between the formation heights of K \textsc{i} and Ca \textsc{ii} wings. Our observations therefore localize the reconnection height to a relatively narrow range, which complements previous studies that probed similar bi-directional signatures with only single chromospheric lines or with non-simultaneous multi-line data; the seamless multi-line coverage achieved by SCIP enables this height localization within a single coherent dataset.

The multi-line capability of SCIP allowed us to probe the height variation of the magnetic field. As shown in Figures \ref{EB1_WFA} and \ref{EB2_WFA}, the derived longitudinal component of the magnetic field decreases with formation height.
This stratification is consistent with the expansion of magnetic flux tubes from the photosphere into the chromosphere. 
Although the WFA is a powerful tool for weak-field regimes, caution is required in the EB kernels where strong velocity gradients exist. 

A key finding of this study is the detection of distinct features of magnetic reconnection in the two events.
In terms of magnetic field structure, SCIP observations clearly reveal the differences in the three-dimensional geometry between the two magnetic reconnection events: opposite polarities were not observed at the formation height of the Ca \textsc{ii} 8542 \AA\ core in Event 1, but they were present in Event 2.
Regarding the heating signatures, Event 1 showed no intensity enhancement in the Ca \textsc{ii} 8542 \AA\ line core, whereas the line core around Event 2 was enhanced by about a factor of two with respect to the quiet Sun.
We note, however, that the core enhancement of Event~2 does not appear as a compact kernel co-spatial with the wing brightenings, unlike the localized EB brightenings observed progressively from the H$\beta$ wings to the line core by \cite{2020A&A...641L...5J}.
Both events are covered by overlying chromospheric fibrils, which appear dark for Event 1 and bright for Event 2 (Figures \ref{example}(c) and (f)), and the core intensity at the event sites therefore cannot be uniquely attributed to local heating at the reconnection site.
For this reason, our interpretation that the reconnecting magnetic structure of Event 2 extends to greater heights relies primarily on the magnetic field topology derived from the WFA, i.e., the persistence of the opposite-polarity structure up to the formation height of the Ca \textsc{ii} 8542 \AA\ core in Event 2 and its absence in Event 1.
The core intensity enhancement of Event 2 is consistent with, though not conclusive of, the associated heating reaching the core formation height. 
The difference in the profiles depending on the heating height of the magnetic reconnection was predicted by \cite{2024ApJ...960...26K}.
We infer that the extension of the current sheet to higher altitudes caused the heating of the upper layers in Event 2.
Arch Filament Systems (AFSs) gradually develop at the site of Event 2 as the flux emergence proceeds (see S. J. González Manrique et al. 2026, ApJL, this issue). It is possible that the magnetic reconnection taking place here is linked to the plasma heating and acceleration observed within the AFS.

Interestingly, although the Ca~\textsc{ii} 8542 \AA\ core in Event~2 shows clear intensity enhancement, no large Doppler shift is detected in the line core itself. A possible explanation is that the observed core emission contains contributions from overlying cold gas, which effectively masks the underlying velocity signatures. Similar Ca \textsc{ii} profiles were reported by \cite{2019A&A...627A.101V}, who demonstrated via non local thermodynamic equilibrium (NLTE) inversions that such events can actually harbor velocity components of 5–20 km/s at chromospheric heights. 
Future work will involve NLTE inversions of the full Stokes profiles acquired by SCIP. Combining this with the high-resolution imaging, we aim to derive the full magnetic vector, velocity field structure and the thermal structure of the bi-directional jets and quantify the energy release rates in the weakly ionized lower solar atmosphere.
\begin{acknowledgments}
\sunriseiii\ is supported by funding from the Max-Planck-Förderstiftung (Max Planck Foundation), NASA under Grants \#80NSSC18K0934 and \#80NSSC24M0024 (“Heliophysics Low Cost Access to Space” program), and the ISAS/JAXA Small Mission-of-Opportunity program and JSPS KAKENHI Grant Numbers JP18H05234 and JP23K25916. This research has received financial support from the European Union’s Horizon 2020 research and innovation program under grant agreement No. 824135 (SOLARNET) and No. 101097844 (WINSUN) from the European Research Council (ERC). It has also been funded by the Deutsches Zentrum für Luft- und Raumfahrt e.V. (DLR, grant no. 50 OO 1608). The Spanish contributions have been funded by the Spanish MCIN/AEI/10.13039/501100011033 under projects RTI2018-096886-B-C5, PID2021-125325OB-C5, and PID2024-156066OB-C5, and from "Center of Excellence Severo Ochoa" awards to IAA-CSIC (SEV-2017-0709, CEX2021-001131-S), all co-funded by "ERDF A way of making Europe". CQN acknowledges support from Grants PID2022-136563NB-I00/10.13039/501100011033, and PID2024-156538NB-I00 and PID2024-156066OB-C55 funded by MCIN/AEI/10.13039/501100011033. This work was carried out by the joint research program of Institute for Space–Earth Environmental Research, Nagoya University. This work is also supported by JSPS KAKENHI grant JP23K13152.
\end{acknowledgments}

\bibliography{reference}{}

\begin{thebibliography}{}
\expandafter\ifx\csname natexlab\endcsname\relax\def\natexlab#1{#1}\fi
\providecommand{\url}[1]{\href{#1}{#1}}
\providecommand{\dodoi}[1]{doi:~\href{http://doi.org/#1}{\nolinkurl{#1}}}
\providecommand{\doeprint}[1]{\href{http://ascl.net/#1}{\nolinkurl{http://ascl.net/#1}}}
\providecommand{\doarXiv}[1]{\href{https://arxiv.org/abs/#1}{\nolinkurl{https://arxiv.org/abs/#1}}}

\bibitem[{P. {Barthol} {et~al.}(2011){Barthol}, {Gandorfer}, {Solanki},
  {Sch{\"u}ssler}, {Chares}, {Curdt}, {Deutsch}, {Feller}, {Germerott},
  {Grauf}, {Heerlein}, {Hirzberger}, {Kolleck}, {Meller}, {M{\"u}ller},
  {Riethm{\"u}ller}, {Tomasch}, {Kn{\"o}lker}, {Lites}, {Card}, {Elmore},
  {Fox}, {Lecinski}, {Nelson}, {Summers}, {Watt}, {Mart{\'\i}nez Pillet},
  {Bonet}, {Schmidt}, {Berkefeld}, {Title}, {Domingo}, {Gasent Blesa}, {del
  Toro Iniesta}, {L{\'o}pez Jim{\'e}nez}, {{\'A}lvarez-Herrero},
  {Sabau-Graziati}, {Widani}, {Haberler}, {H{\"a}rtel}, {Kampf}, {Levin},
  {P{\'e}rez Grande}, {Sanz-Andr{\'e}s}, \& {Schmidt}}]{2011SoPh..268....1B}
{Barthol}, P., {Gandorfer}, A., {Solanki}, S.~K., {et~al.} 2011,
  \bibinfo{title}{{The Sunrise Mission},} \solphys, 268, 1,
  \dodoi{10.1007/s11207-010-9662-9}

\bibitem[{T. {Berkefeld} {et~al.}(2026){Berkefeld}, {Bell}, {Volkmer},
  {Heidecke}, {Preis}, {Sonner}, {Nakai}, {Korpi-Lagg}, {Gandorfer}, {Solanki},
  {del Toro Iniesta}, {Katsukawa}, {Bernasconi}, {Feller}, {Riethm{\"u}ller},
  {{\'A}lvarez-Herrero}, {Kubo}, {Mart{\'\i}nez Pillet}, {Smitha}, {Orozco
  Su{\'a}rez}, {Grauf}, \& {Carpenter}}]{2026SoPh..301...57B}
{Berkefeld}, T., {Bell}, A., {Volkmer}, R., {et~al.} 2026,
  \bibinfo{title}{{SUNRISE III: The Wavefront Correction System},} \solphys,
  301, 57, \dodoi{10.1007/s11207-026-02649-7}

\bibitem[{P. {Bernasconi} {et~al.}(2025){Bernasconi}, {Carpenter}, {Eaton},
  {Schulze}, {Carkhuff}, {Palo}, {Young}, {Raouafi}, {Vourlidas}, {Coker},
  {Solanki}, {Korpi-Lagg}, {Gandorfer}, {Feller}, {Riethm{\"u}ller}, {Smitha},
  {Grauf}, {del Toro Iniesta}, {Orozco Su{\'a}rez}, {Katsukawa}, {Kubo},
  {Berkefeld}, {Bell}, {{\'A}lvarez-Herrero}, \& {Mart{\'\i}nez
  Pillet}}]{2025SoPh..300..112B}
{Bernasconi}, P., {Carpenter}, M., {Eaton}, H., {et~al.} 2025,
  \bibinfo{title}{{The Gondola for the SUNRISE III Balloon-Borne Solar
  Observatory},} \solphys, 300, 112, \dodoi{10.1007/s11207-025-02524-x}

\bibitem[{R. {Centeno}(2018){Centeno}}]{2018ApJ...866...89C}
{Centeno}, R. 2018, \bibinfo{title}{{On the Weak Field Approximation for Ca
  8542 {\r{A}}},} \apj, 866, 89, \dodoi{10.3847/1538-4357/aae087}

\bibitem[{F. {Ellerman}(1917){Ellerman}}]{1917ApJ....46..298E}
{Ellerman}, F. 1917, \bibinfo{title}{{Solar Hydrogen ``bombs''},} \apj, 46,
  298, \dodoi{10.1086/142366}

\bibitem[{J.~M. {Fontenla} {et~al.}(1993){Fontenla}, {Avrett}, \&
  {Loeser}}]{1993ApJ...406..319F}
{Fontenla}, J.~M., {Avrett}, E.~H., \& {Loeser}, R. 1993,
  \bibinfo{title}{{Energy Balance in the Solar Transition Region. III. Helium
  Emission in Hydrostatic, Constant-Abundance Models with Diffusion},} \apj,
  406, 319, \dodoi{10.1086/172443}

\bibitem[{G.~A. {Gary}(2001){Gary}}]{2001SoPh..203...71G}
{Gary}, G.~A. 2001, \bibinfo{title}{{Plasma Beta above a Solar Active Region:
  Rethinking the Paradigm},} \solphys, 203, 71, \dodoi{10.1023/A:1012722021820}

\bibitem[{M.~K. {Georgoulis} {et~al.}(2002){Georgoulis}, {Rust}, {Bernasconi},
  \& {Schmieder}}]{2002ApJ...575..506G}
{Georgoulis}, M.~K., {Rust}, D.~M., {Bernasconi}, P.~N., \& {Schmieder}, B.
  2002, \bibinfo{title}{{Statistics, Morphology, and Energetics of Ellerman
  Bombs},} \apj, 575, 506, \dodoi{10.1086/341195}

\bibitem[{J. {Jefferies} {et~al.}(1989){Jefferies}, {Lites}, \&
  {Skumanich}}]{1989ApJ...343..920J}
{Jefferies}, J., {Lites}, B.~W., \& {Skumanich}, A. 1989,
  \bibinfo{title}{{Transfer of Line Radiation in a Magnetic Field},} \apj, 343,
  920, \dodoi{10.1086/167762}

\bibitem[{J. {Joshi} {et~al.}(2020){Joshi}, {Rouppe van der Voort}, \& {de la
  Cruz Rodr{\'\i}guez}}]{2020A&A...641L...5J}
{Joshi}, J., {Rouppe van der Voort}, L. H.~M., \& {de la Cruz Rodr{\'\i}guez},
  J. 2020, \bibinfo{title}{{Signatures of ubiquitous magnetic reconnection in
  the lower solar atmosphere},} \aap, 641, L5,
  \dodoi{10.1051/0004-6361/202038769}

\bibitem[{Y. {Katsukawa} {et~al.}(2020){Katsukawa}, {del Toro Iniesta},
  {Solanki}, {Kubo}, {Hara}, {Shimizu}, {Oba}, {Kawabata}, {Tsuzuki},
  {Uraguchi}, {Nodomi}, {Shinoda}, {Tamura}, {Suematsu}, {Ishikawa}, {Kano},
  {Matsumoto}, {Ichimoto}, {Nagata}, {Quintero Noda}, {Anan}, {Orozco
  Su{\'a}rez}, {Balaguer Jim{\'e}nez}, {L{\'o}pez Jim{\'e}nez}, {Cobos
  Carrascosa}, {Feller}, {Riethmueller}, {Gandorfer}, \&
  {Lagg}}]{2020SPIE11447E..0YK}
{Katsukawa}, Y., {del Toro Iniesta}, J.~C., {Solanki}, S.~K., {et~al.} 2020,
  \bibinfo{title}{{Sunrise Chromospheric Infrared SpectroPolarimeter (SCIP) for
  sunrise III: system design and capability},} in Society of Photo-Optical
  Instrumentation Engineers (SPIE) Conference Series, Vol. 11447, Society of
  Photo-Optical Instrumentation Engineers (SPIE) Conference Series, 114470Y,
  \dodoi{10.1117/12.2561223}

\bibitem[{Y. {Katsukawa} {et~al.}(2026){Katsukawa}, {del Toro Iniesta},
  {Solanki}, {Kubo}, {Hara}, {Shimizu}, {Oba}, {Kawabata}, {Tsuzuki},
  {Uraguchi}, {Shinoda}, {Tamura}, {Suematsu}, {Matsumoto}, {Ishikawa},
  {Naito}, {Ichimoto}, {Nagata}, {Anan}, {Orozco Su{\'a}rez}, {Sanchis
  Kilders}, {Balaguer Jim{\'e}nez}, {L{\'o}pez Jim{\'e}nez}, {Quintero Noda},
  {{\'A}lvarez Garc{\'\i}a}, {Ramos M{\'a}s}, {Cobos Carrascosa}, {Labrousse},
  {Aparicio del Moral}, {S{\'a}nchez G{\'o}mez}, {Hern{\'a}ndez Exp{\'o}sito},
  {Bail{\'o}n Mart{\'\i}nez}, {Morales Fern{\'a}ndez}, {Moreno Mantas},
  {Tobaruela}, {Bustamante}, {Bail{\'e}n}, {Blanco Rodr{\'\i}guez}, {Gasent
  Blesa}, {Rodr{\'\i}guez Mart{\'\i}nez}, {Ferreres}, {Gilabert Palmer},
  {Piqueras Carre{\~n}o}, {P{\'e}rez Grande}, {Torralbo},
  {{\'A}lvarez-Herrero}, {Korpi-Lagg}, {Gandorfer}, {Berkefeld}, {Bernasconi},
  {Feller}, {Riethm{\"u}ller}, {Smitha}, {Mart{\'\i}nez Pillet}, {Grauf},
  {Bell}, \& {Carpenter}}]{2026SoPh..301...99K}
{Katsukawa}, Y., {del Toro Iniesta}, J.~C., {Solanki}, S.~K., {et~al.} 2026,
  \bibinfo{title}{{The Sunrise Chromospheric Infrared Spectro-Polarimeter SCIP:
  An Instrument for SUNRISE III},} \solphys, 301, 99,
  \dodoi{10.1007/s11207-026-02696-0}

\bibitem[{Y. {Kawabata} {et~al.}(2024){Kawabata}, {Quintero Noda}, {Katsukawa},
  {Kubo}, {Matsumoto}, \& {Oba}}]{2024ApJ...960...26K}
{Kawabata}, Y., {Quintero Noda}, C., {Katsukawa}, Y., {et~al.} 2024,
  \bibinfo{title}{{Multiline Stokes Synthesis of Ellerman Bombs: Obtaining
  Seamless Information from Photosphere to Chromosphere},} \apj, 960, 26,
  \dodoi{10.3847/1538-4357/acf9fc}

\bibitem[{R. {Kitai}(1983){Kitai}}]{1983SoPh...87..135K}
{Kitai}, R. 1983, \bibinfo{title}{{On the mass motions and the atmospheric
  states of moustaches.},} \solphys, 87, 135, \dodoi{10.1007/BF00151165}

\bibitem[{A. {Korpi-Lagg} {et~al.}(2025){Korpi-Lagg}, {Gandorfer}, {Solanki},
  {del Toro Iniesta}, {Katsukawa}, {Bernasconi}, {Berkefeld}, {Feller},
  {Riethm{\"u}ller}, {{\'A}lvarez-Herrero}, {Kubo}, {Mart{\'\i}nez Pillet},
  {Smitha}, {Orozco Su{\'a}rez}, {Grauf}, {Carpenter}, {Bell},
  {{\'A}lvarez-Alonso}, {{\'A}lvarez Garc{\'\i}a}, {Aparicio del Moral},
  {Ati{\'e}nzar}, {Ayoub}, {Bail{\'e}n}, {Bail{\'o}n Mart{\'\i}nez}, {Balaguer
  Jim{\'e}nez}, {Barthol}, {Bayon Laguna}, {Bellot Rubio}, {Bergmann}, {Blanco
  Rodr{\'\i}guez}, {Bochmann}, {Borrero}, {Campos-Jara}, {Castellanos
  Dur{\'a}n}, {Cebollero}, {Conde Rodr{\'\i}guez}, {Deutsch}, {Eaton},
  {Fern{\'a}ndez-Medina}, {Fernandez-Rico}, {Ferreres}, {Garc{\'\i}a},
  {Garc{\'\i}a Alarcia}, {Garc{\'\i}a Parejo}, {Garranzo-Garc{\'\i}a}, {Gasent
  Blesa}, {Gerber}, {Germerott}, {Gilabert Palmer}, {Gizon}, {G{\'o}mez
  S{\'a}nchez-Tirado}, {Gonz{\'a}lez-B{\'a}rcena}, {Gonzalo Melchor},
  {Goodyear}, {Hara}, {Harnes}, {Heerlein}, {Heidecke}, {Heinrichs},
  {Hern{\'a}ndez Exp{\'o}sito}, {Hirzberger}, {Hoelken}, {Hyun}, {Iglesias},
  {Ishikawa}, {Jeon}, {Kawabata}, {Kolleck}, {Laguna}, {Lomas}, {L{\'o}pez
  Jim{\'e}nez}, {Manzano}, {Matsumoto}, {Mayo Turrado}, {Meierdierks},
  {Meining}, {Monecke}, {Morales-Fern{\'a}ndez}, {Moreno Mantas}, {Moreno
  Vacas}, {M{\"u}ller}, {M{\"u}ller}, {Naito}, {Nakai}, {N{\'u}{\~n}ez Peral},
  {Oba}, {Palo}, {P{\'e}rez-Grande}, {Piqueras Carre{\~n}o}, {Preis},
  {Przybylski}, {Quintero Noda}, {Ramanath}, {Ramos M{\'a}s}, {Raouafi},
  {Rivas-Mart{\'\i}nez}, {Rodr{\'\i}guez Mart{\'\i}nez}, {Rodr{\'\i}guez
  Valido}, {Ruiz Cobo}, {S{\'a}nchez Rodr{\'\i}guez}, {Sanchez Toledo},
  {S{\'a}nchez G{\'o}mez}, {Sanchis Kilders}, {Sant}, {Santamarina Guerrero},
  {Schulze}, {Shimizu}, {Silva-L{\'o}pez}, {Singh}, {Siu-Tapia}, {Sonner},
  {Staub}, {Strecker}, {Tobaruela}, {Torralbo}, {Tritschler}, {Tsuzuki},
  {Uraguchi}, {Volkmer}, {Vourlidas}, {Vukadinovi{\'c}}, {Werner}, \&
  {Zerr}}]{2025SoPh..300...75K}
{Korpi-Lagg}, A., {Gandorfer}, A., {Solanki}, S.~K., {et~al.} 2025,
  \bibinfo{title}{{SUNRISE III: Overview of Observatory and Instruments},}
  \solphys, 300, 75, \dodoi{10.1007/s11207-025-02485-1}

\bibitem[{T. {Matsumoto} {et~al.}(2023){Matsumoto}, {Kawabata}, {Katsukawa},
  {Iijima}, \& {Quintero Noda}}]{2023MNRAS.523..974M}
{Matsumoto}, T., {Kawabata}, Y., {Katsukawa}, Y., {Iijima}, H., \& {Quintero
  Noda}, C. 2023, \bibinfo{title}{{Synthesis of infrared Stokes spectra in an
  evolving solar chromospheric jet},} \mnras, 523, 974,
  \dodoi{10.1093/mnras/stad1509}

\bibitem[{T. {Matsumoto} {et~al.}(2008){Matsumoto}, {Kitai}, {Shibata},
  {Otsuji}, {Naruse}, {Shiota}, \& {Takasaki}}]{2008PASJ...60...95M}
{Matsumoto}, T., {Kitai}, R., {Shibata}, K., {et~al.} 2008,
  \bibinfo{title}{{Height Dependence of Gas Flows in an Ellerman Bomb},} \pasj,
  60, 95, \dodoi{10.1093/pasj/60.1.95}

\bibitem[{R. {Morosin} {et~al.}(2020){Morosin}, {de la Cruz Rodr{\'\i}guez},
  {Vissers}, \& {Yadav}}]{2020A&A...642A.210M}
{Morosin}, R., {de la Cruz Rodr{\'\i}guez}, J., {Vissers}, G. J.~M., \&
  {Yadav}, R. 2020, \bibinfo{title}{{Stratification of canopy magnetic fields
  in a plage region. Constraints from a spatially-regularized weak-field
  approximation method},} \aap, 642, A210, \dodoi{10.1051/0004-6361/202038754}

\bibitem[{A. {Ortiz} {et~al.}(2020){Ortiz}, {Hansteen}, {N{\'o}brega-Siverio},
  \& {van der Voort}}]{2020A&A...633A..58O}
{Ortiz}, A., {Hansteen}, V.~H., {N{\'o}brega-Siverio}, D., \& {van der Voort},
  L.~R. 2020, \bibinfo{title}{{Ellerman bombs and UV bursts: reconnection at
  different atmospheric layers},} \aap, 633, A58,
  \dodoi{10.1051/0004-6361/201936574}

\bibitem[{H. {Peter} {et~al.}(2019){Peter}, {Huang}, {Chitta}, \&
  {Young}}]{2019A&A...628A...8P}
{Peter}, H., {Huang}, Y.~M., {Chitta}, L.~P., \& {Young}, P.~R. 2019,
  \bibinfo{title}{{Plasmoid-mediated reconnection in solar UV bursts},} \aap,
  628, A8, \dodoi{10.1051/0004-6361/201935820}

\bibitem[{H. {Peter} {et~al.}(2014){Peter}, {Tian}, {Curdt}, {Schmit}, {Innes},
  {De Pontieu}, {Lemen}, {Title}, {Boerner}, {Hurlburt}, {Tarbell}, {Wuelser},
  {Mart{\'\i}nez-Sykora}, {Kleint}, {Golub}, {McKillop}, {Reeves}, {Saar},
  {Testa}, {Kankelborg}, {Jaeggli}, {Carlsson}, \&
  {Hansteen}}]{2014Sci...346C.315P}
{Peter}, H., {Tian}, H., {Curdt}, W., {et~al.} 2014, \bibinfo{title}{{Hot
  explosions in the cool atmosphere of the Sun},} Science, 346, 1255726,
  \dodoi{10.1126/science.1255726}

\bibitem[{C. {Quintero Noda} {et~al.}(2016){Quintero Noda}, {Shimizu}, {de la
  Cruz Rodr{\'\i}guez}, {Katsukawa}, {Ichimoto}, {Anan}, \&
  {Suematsu}}]{2016MNRAS.459.3363Q}
{Quintero Noda}, C., {Shimizu}, T., {de la Cruz Rodr{\'\i}guez}, J., {et~al.}
  2016, \bibinfo{title}{{Spectropolarimetric capabilities of Ca II 8542 {\r{A}}
  line},} \mnras, 459, 3363, \dodoi{10.1093/mnras/stw867}

\bibitem[{C. {Quintero Noda} {et~al.}(2017{\natexlab{a}}){Quintero Noda},
  {Shimizu}, {Katsukawa}, {de la Cruz Rodr{\'\i}guez}, {Carlsson}, {Anan},
  {Oba}, {Ichimoto}, \& {Suematsu}}]{2017MNRAS.464.4534Q}
{Quintero Noda}, C., {Shimizu}, T., {Katsukawa}, Y., {et~al.}
  2017{\natexlab{a}}, \bibinfo{title}{{Chromospheric polarimetry through
  multiline observations of the 850-nm spectral region},} \mnras, 464, 4534,
  \dodoi{10.1093/mnras/stw2738}

\bibitem[{C. {Quintero Noda} {et~al.}(2017{\natexlab{b}}){Quintero Noda},
  {Kato}, {Katsukawa}, {Oba}, {de la Cruz Rodr{\'\i}guez}, {Carlsson},
  {Shimizu}, {Orozco Su{\'a}rez}, {Ruiz Cobo}, {Kubo}, {Anan}, {Ichimoto}, \&
  {Suematsu}}]{2017MNRAS.472..727Q}
{Quintero Noda}, C., {Kato}, Y., {Katsukawa}, Y., {et~al.} 2017{\natexlab{b}},
  \bibinfo{title}{{Chromospheric polarimetry through multiline observations of
  the 850-nm spectral region - II. A magnetic flux tube scenario},} \mnras,
  472, 727, \dodoi{10.1093/mnras/stx2022}

\bibitem[{C. {Quintero Noda} {et~al.}(2019){Quintero Noda}, {Iijima},
  {Katsukawa}, {Shimizu}, {Carlsson}, {de la Cruz Rodr{\'\i}guez}, {Ruiz Cobo},
  {Orozco Su{\'a}rez}, {Oba}, {Anan}, {Kubo}, {Kawabata}, {Ichimoto}, \&
  {Suematsu}}]{2019MNRAS.486.4203Q}
{Quintero Noda}, C., {Iijima}, H., {Katsukawa}, Y., {et~al.} 2019,
  \bibinfo{title}{{Chromospheric polarimetry through multiline observations of
  the 850 nm spectral region III: Chromospheric jets driven by twisted magnetic
  fields},} \mnras, 486, 4203, \dodoi{10.1093/mnras/stz1124}

\bibitem[{L.~H.~M. {Rouppe van der Voort} {et~al.}(2016){Rouppe van der Voort},
  {Rutten}, \& {Vissers}}]{2016A&A...592A.100R}
{Rouppe van der Voort}, L. H.~M., {Rutten}, R.~J., \& {Vissers}, G. J.~M. 2016,
  \bibinfo{title}{{Reconnection brightenings in the quiet solar photosphere},}
  \aap, 592, A100, \dodoi{10.1051/0004-6361/201628889}

\bibitem[{R.~J. {Rutten} {et~al.}(2013){Rutten}, {Vissers}, {Rouppe van der
  Voort}, {S{\"u}tterlin}, \& {Vitas}}]{2013JPhCS.440a2007R}
{Rutten}, R.~J., {Vissers}, G. J.~M., {Rouppe van der Voort}, L. H.~M.,
  {S{\"u}tterlin}, P., \& {Vitas}, N. 2013, \bibinfo{title}{{Ellerman bombs:
  fallacies, fads, usage},} in Journal of Physics Conference Series, Vol. 440,
  Journal of Physics Conference Series, 012007,
  \dodoi{10.1088/1742-6596/440/1/012007}

\bibitem[{P.~D. {Smith} \& J.~I. {Sakai}(2008){Smith} \&
  {Sakai}}]{2008A&A...486..569S}
{Smith}, P.~D., \& {Sakai}, J.~I. 2008, \bibinfo{title}{{Chromospheric magnetic
  reconnection: two-fluid simulations of coalescing current loops},} \aap, 486,
  569, \dodoi{10.1051/0004-6361:200809624}

\bibitem[{S.~K. {Solanki} {et~al.}(2010){Solanki}, {Barthol}, {Danilovic},
  {Feller}, {Gandorfer}, {Hirzberger}, {Riethm{\"u}ller}, {Sch{\"u}ssler},
  {Bonet}, {Mart{\'\i}nez Pillet}, {del Toro Iniesta}, {Domingo}, {Palacios},
  {Kn{\"o}lker}, {Bello Gonz{\'a}lez}, {Berkefeld}, {Franz}, {Schmidt}, \&
  {Title}}]{2010ApJ...723L.127S}
{Solanki}, S.~K., {Barthol}, P., {Danilovic}, S., {et~al.} 2010,
  \bibinfo{title}{{SUNRISE: Instrument, Mission, Data, and First Results},}
  \apjl, 723, L127, \dodoi{10.1088/2041-8205/723/2/L127}

\bibitem[{S.~K. {Solanki} {et~al.}(2017){Solanki}, {Riethm{\"u}ller},
  {Barthol}, {Danilovic}, {Deutsch}, {Doerr}, {Feller}, {Gandorfer},
  {Germerott}, {Gizon}, {Grauf}, {Heerlein}, {Hirzberger}, {Kolleck}, {Lagg},
  {Meller}, {Tomasch}, {van Noort}, {Blanco Rodr{\'\i}guez}, {Gasent Blesa},
  {Balaguer Jim{\'e}nez}, {Del Toro Iniesta}, {L{\'o}pez Jim{\'e}nez}, {Orozco
  Suarez}, {Berkefeld}, {Halbgewachs}, {Schmidt}, {{\'A}lvarez-Herrero},
  {Sabau-Graziati}, {P{\'e}rez Grande}, {Mart{\'\i}nez Pillet}, {Card},
  {Centeno}, {Kn{\"o}lker}, \& {Lecinski}}]{2017ApJS..229....2S}
{Solanki}, S.~K., {Riethm{\"u}ller}, T.~L., {Barthol}, P., {et~al.} 2017,
  \bibinfo{title}{{The Second Flight of the Sunrise Balloon-borne Solar
  Observatory: Overview of Instrument Updates, the Flight, the Data, and First
  Results},} \apjs, 229, 2, \dodoi{10.3847/1538-4365/229/1/2}

\bibitem[{S.~K. {Solanki} {et~al.}(2026){Solanki}, {Smitha}, {Lagg},
  {Gandorfer}, {del Toro Iniesta}, {Katsukawa}, {Bernasconi}, {Berkefeld},
  {Feller}, {Riethm{\"u}ller}, {{\'A}lvarez-Herrero}, {Kubo}, {Orozco
  Su{\'a}rez}, {Grauf}, {Carpenter}, {Bell}, {Mart{\'\i}nez Pillet}, {Gizon},
  {Bail{\'e}n}, {Blanco Rodr{\'\i}guez}, {Sebasti{\'a}n Castellanos Dur{\'a}n},
  {Harnes}, {Hoelken}, {Iglesias}, {Ishikawa}, {Kawabata}, {Matsumoto}, {Oba},
  {Singh}, {Siu-Tapia}, {Strecker}, {Vukadinovi{\'c}}, {van Noort}, {Balaguer
  Jim{\'e}nez}, {Sanchis Kilders}, {Torralbo}, {Kuckein}, {Hara}, {Shimizu},
  {Volkmer}, {Preis}, {Raouafi}, {Vourlidas}, {Hirzberger}, {Deutsch},
  {Germerott}, {Heerlein}, {Kolleck}, {{\'A}lvarez Garc{\'\i}a}, {L{\'o}pez
  Jim{\'e}nez}, {Bellot Rubio}, {Morales-Fern{\'a}ndez}, {Jes{\'u}s Moreno
  Mantas}, {Aparicio del Moral}, {S{\'a}nchez G{\'o}mez}, {Bail{\'o}n
  Mart{\'\i}nez}, {Santamarina Guerrero}, {Hern{\'a}ndez Exp{\'o}sito},
  {Tobaruela}, {Gasent Blesa}, {Schulze}, {Eaton}, {Palo}, {Ayoub}, {Naito},
  {Quintero Noda}, {Uraguchi}, {Tsuzuki}, \& {Piqueras
  Carre{\~n}o}}]{2026arXiv260607989S}
{Solanki}, S.~K., {Smitha}, H.~N., {Lagg}, A., {et~al.} 2026,
  \bibinfo{title}{{Sunrise III: Instrument, mission, data, and first results},}
  arXiv e-prints, arXiv:2606.07989.
\newblock \doarXiv{2606.07989}

\bibitem[{W. {Unno}(1956){Unno}}]{1956PASJ....8..108U}
{Unno}, W. 1956, \bibinfo{title}{{Line Formation of a Normal Zeeman Triplet},}
  \pasj, 8, 108, \dodoi{10.1093/pasj/8.3-4.108}

\bibitem[{G.~J.~M. {Vissers} {et~al.}(2019){Vissers}, {de la Cruz
  Rodr{\'\i}guez}, {Libbrecht}, {Rouppe van der Voort}, {Scharmer}, \&
  {Carlsson}}]{2019A&A...627A.101V}
{Vissers}, G.~J.~M., {de la Cruz Rodr{\'\i}guez}, J., {Libbrecht}, T., {et~al.}
  2019, \bibinfo{title}{{Dissecting bombs and bursts: non-LTE inversions of
  low-atmosphere reconnection in SST and IRIS observations},} \aap, 627, A101,
  \dodoi{10.1051/0004-6361/201833560}

\bibitem[{G.~J.~M. {Vissers} {et~al.}(2013){Vissers}, {Rouppe van der Voort},
  \& {Rutten}}]{2013ApJ...774...32V}
{Vissers}, G. J.~M., {Rouppe van der Voort}, L. H.~M., \& {Rutten}, R.~J. 2013,
  \bibinfo{title}{{Ellerman Bombs at High Resolution. II. Triggering,
  Visibility, and Effect on Upper Atmosphere},} \apj, 774, 32,
  \dodoi{10.1088/0004-637X/774/1/32}

\bibitem[{G.~J.~M. {Vissers} {et~al.}(2015){Vissers}, {Rouppe van der Voort},
  {Rutten}, {Carlsson}, \& {De Pontieu}}]{2015ApJ...812...11V}
{Vissers}, G.~J.~M., {Rouppe van der Voort}, L.~H.~M., {Rutten}, R.~J.,
  {Carlsson}, M., \& {De Pontieu}, B. 2015, \bibinfo{title}{{Ellerman Bombs at
  High Resolution. III. Simultaneous Observations with IRIS and SST},} \apj,
  812, 11, \dodoi{10.1088/0004-637X/812/1/11}

\bibitem[{H. {Watanabe} {et~al.}(2011){Watanabe}, {Vissers}, {Kitai}, {Rouppe
  van der Voort}, \& {Rutten}}]{2011ApJ...736...71W}
{Watanabe}, H., {Vissers}, G., {Kitai}, R., {Rouppe van der Voort}, L., \&
  {Rutten}, R.~J. 2011, \bibinfo{title}{{Ellerman Bombs at High Resolution. I.
  Morphological Evidence for Photospheric Reconnection},} \apj, 736, 71,
  \dodoi{10.1088/0004-637X/736/1/71}

\end{thebibliography}
\bibliographystyle{aasjournalv7}

\end{document}